# Strain-driven InAs island growth on top of GaAs(111) nanopillars

T. Riedl,[*] V. S. Kunnathully, A. Trapp, T. Langer, D. Reuter, and J. K. N. Lindner

*Department of Physics and Center for Optoelectronics and Photonics Paderborn (CeOPP),
Paderborn University, Warburger Straße 100, 33098 Paderborn, Germany*

[*]Corresponding author: thomas.riedl@uni-paderborn.de

We analyze the shape and position of heteroepitaxial InAs islands on the top face of cylindrical GaAs(111)A nanopillars experimentally and theoretically. Catalyst-free molecular beam epitaxial growth of InAs at low temperatures on GaAs nanopillars results in InAs islands with diameters < 30 nm exhibiting predominantly rounded triangular in-plane shapes. The islands show a tendency to grow at positions displaced from the center towards the pillar edge. Atomistic molecular statics simulations evidence that triangular-prismatic islands centered to the pillar axis with diameters smaller than that of the nanopillars are energetically favored. Moreover, we reveal the existence of minimum-energy states for off-axis island positions, in agreement with the experiment. These findings are interpreted by evaluating the spatial strain distributions and the number of broken bonds of surface atoms as a measure for the surface energy. The preferred off-axis island positions can be understood in terms of an increased compliancy of the GaAs nanopillar beneath the island because of the vicinity of free surfaces, leading to a reduction of strain energy. The influence of surface steps on the energy of the system is addressed as well.

## I. INTRODUCTION

The size, shape and position of semiconductor nanostructures have proven to be essential for their application in optoelectronics. Regarding heteroepitaxially grown quantum dots (QDs), position control has been achieved by means of pits prepatterned on the substrate surface [1]. Alternatively, the top face of nanopillars or wires either etched into or grown onto the substrate can be exploited as growth area for single QDs or disk-like nanolayers [2]. Particular attention has been paid to QD-in-wire heterostructures, which are promising for realizing tunneling devices [3], light emitting devices [4] and single-photon sources [5]. It has been shown theoretically that in catalyst-free epitaxy the diameter of the QD or island grown on top of a nanowire can be equal or smaller than that of the nanowire substrate [6]. This results from the trade-off between misfit induced strain energy and surface energy, which elicits an energy minimum for a specific heterolayer or island diameter, dependent on lattice misfit, nanowire diameter, surface energies and layer thickness. Experimental evidence for such reduced diameter islands is reported in the case of ternary $In_{1-x}Ga_xN$ axially grown on top of GaN nanowires [7], where the $In_{1-x}Ga_xN$ adopts the shape of a column surrounded by a GaN shell. The $In_{1-x}Ga_xN$ morphology is also influenced by the growth conditions, notably the In/Ga flux ratio, because it determines the formation or non-formation of an In wetting layer and thus the surface energy [8]. If several In$_{1-x}$Ga$_x$N quantum disks are stacked along the growth direction in GaN nanowires, the disk diameter is seen to increase with increasing disk number in the stack, which is attributed to a vertical, shape-dependent strain interaction between the disks [9].

Apart from the formation of smaller diameter islands on top of nanowires, it is well known that QDs of larger lattice parameter than the substrate nucleate preferably at convex edges of the substrate [10]. At these sites, the wetting layer is less compressively strained due to the increased elastic lattice relaxation, which reduces the chemical potential of the surface [11]. There has not yet been much study on how this effect modifies the strain energy – surface energy interplay in the case of QDs growing on top of nanopillars.

In the present study we analyze the morphology of InAs islands grown on GaAs(111) nanopillars patterned into the substrate experimentally and theoretically. InAs QDs are attractive for achieving infrared emission for optoelectronic and telecommunication applications, e.g. QD lasers [12], QD infrared photodetectors [13], single photon sources [14] and solar cells [15]. Moreover, the growth on the nanopillars enables the fabrication of InAs QDs on (111)-oriented GaAs substrate, which is not attainable on planar GaAs(001). In addition to paying attention to the island aspect ratio, we also elucidate the dependence of the system energy on the *radial* island position on the nanopillar top face. Atomistic calculations based on empirical potentials are employed, which allow for a sufficiently accurate description of total energy, strain distribution, and surface energy.

## II. EXPERIMENTAL METHODS

GaAs(111)A wafers were nanopillar-patterned using nanosphere lithography and reactive ion etching. First monolayers and double layers of polystyrene spheres with a diameter of 220 nm were deposited by means of the doctor-blade technique on the hydrophilized substrate surface. By deposition of Ni and removal of polystyrene spheres, Ni hard masks were fabricated from which nanopillars were formed by anisotropic SiCl$_4$ reactive ion etching. The pillar height amounted to 80–90 nm, the diameter 20–45 nm. Residual Ni and surface oxides were dissolved wet-chemically in diluted H$_2$SO$_4$ and HF solution, respectively. Details on the patterning process can be found in a recent paper [16].

Heteroepitaxial growth of InAs on the nanopillar-patterned GaAs surface was performed by solid source molecular beam epitaxy after atomic H cleaning of the patterned substrate. In order to obtain InAs growth on the nanopillars, a low growth temperature of 150 °C at a rate of 0.011 nm/s under As-rich conditions (V/III ratio ~400) was chosen [16]. The nominally deposited InAs thickness was 15 nm.

Morphological and structural characterization of the heteroepitaxially overgrown substrates was performed by high-resolution scanning electron microscopy (SEM) and transmission electron microscopy (TEM) imaging. As instruments a Raith Pioneer field-emission SEM operated at 15 kV and a JEOL JEM-ARM200F TEM operated at 200 kV were used. TEM cross-sectional specimens were prepared by mechanical grinding followed by dimpling and ion polishing using a Gatan PIPS Model 691.

## III. THEORETICAL CALCULATIONS

For analyzing the strain and energy of heteroepitaxial InAs islands on top of GaAs(111)A nanopillars in dependence of the island aspect ratio and position on the pillar top face, atomistic molecular statics calculations based on the Tersoff potential [17] were performed. Although the experimentally prepared GaAs nanopillars mostly had rounded triangular in-plane shapes, a perfectly cylindrical pillar geometry was considered in the simulations, for the sake of simplicity. The diameter of the GaAs pillar amounted to 25 nm, whereas its length was chosen much larger than the InAs thickness. On the GaAs (111) top face, a one monolayer thin InAs wetting layer and for the InAs island a triangular prismatic shape with $\{11\bar{2}\}$ sidewall facets are considered (Fig. 1), as this corresponds approximately to the experimental observations. While the island width was varied, its height was adjusted in order to keep the number of atoms in the island constant. For the purpose of analyzing the energetics during the initial growth stage the number of In and As atoms (including the wetting layer) was chosen such as to yield a coverage of approximately 2.7 monolayers on the entire pillar (111)A top surface, respectively. Consequently, no misfit dislocation was introduced, since plastic relaxation occurs for larger deposited thicknesses but is assumed not to change the island position. The position of the InAs island was varied along the $\pm[11\bar{2}]$ directions, i.e. perpendicular to one of the island sidewalls, in order to study the two limiting cases with either the triangle edge or the triangle tip approaching the edge of the GaAs pillar top face. At first, atom coordinates were generated by self-written script programs in the DigitalMicrograph software [18]. Then, the structures were iteratively relaxed by minimizing their total energy with the conjugate gradient method in the LAMMPS software [19]. In the Tersoff potential approach, the total energy is computed as the sum of repulsive and attractive interaction energies of atom triples taking the effect of atomic coordination on bond strength into account. The potential parametrization of Hammerschmidt et al. was chosen, because it allows for an accurate description of GaAs and InAs bulk and surfaces [20]. This parametrization yields an equilibrium lattice parameter of 5.654 Å for GaAs and 6.058 Å for InAs. Iterations were stopped once the relative energy difference between successive steps decreased below $10^{-9}$. Strain distributions and atomic coordination of surface atoms were extracted with self-written DigitalMicrograph scripts. The strain is calculated as atomistic strain by evaluating the distances between atoms in the relaxed, strained state and comparing with those in the unstrained state of the respective material. For estimation of the strain energy a Young's modulus of 85.5 and 51.4 GPa, and a Poisson ratio of 0.31 and 0.35 are used for GaAs [21] and InAs [22], respectively.

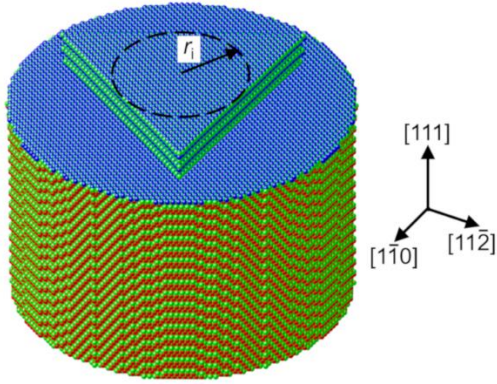

FIG. 1. Atomistic representation (orthographic projection) of one of the analyzed InAs islands (incircle radius $r_i$) on a GaAs nanopillar (diameter 25 nm) covered with one monolayer InAs wetting layer on the (111)A surface after molecular statics relaxation. Red beads represent Ga, blue ones In, and green ones As atoms.

## IV. RESULTS AND DISCUSSION

Experimentally we observe that for the employed growth conditions the deposited InAs grows in the form of islands on the (111)A top face of the GaAs nanopillars (Fig. 2) with their characteristic in-plane dimensions typically smaller than the pillar diameter. This is a clear qualitative confirmation of the predictions of a previous theoretical study which found that islanding is energetically favored over disk-like growth for not too small nanopillar diameters and misfit [6]. As visible in the topview SEM image (Fig. 2(a)) the islands mostly have rounded triangular in-plane shapes with major edges parallel to $\langle 1\bar{1}0\rangle$ leading to edge normals parallel to $\langle 11\bar{2}\rangle$ directions. Both sets of edge normals, [$11\bar{2}$], [$1\bar{2}1$], [$\bar{2}11$] and [$\bar{1}\bar{1}2$], [$\bar{1}2\bar{1}$], [$2\bar{1}\bar{1}$] are present. Obviously, there is less InAs than the nominally deposited 15 nm on top of the pillars, which can be attributed to In adatom diffusion towards the concave edges at the pillar base acting as strong adatom sinks and subsequent crystal growth. Regarding the island position, some are located close to the nanopillar axis (Fig. 2(a) and 2(b)), while others occur at the edge of the nanopillar top face (Fig. 2(a) and 2(c)).

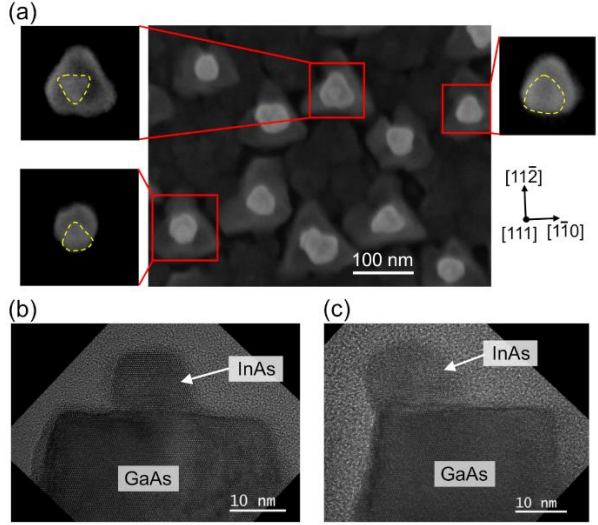

FIG. 2. Representative (a) topview SEM image and (b,c) cross-sectional TEM images in [$1\bar{1}0$] zone axis of nanopillar-patterned GaAs(111)A overgrown with nominally 15 nm InAs. The pillar pattern has been fabricated using nanosphere lithography, which results in the honeycomb arrangement of pillars. In (b) and (c) the structure is embedded in the epoxy used for TEM preparation.

In order to understand the observed morphology, i.e. the aspect ratio (height/diameter) and the position of the InAs islands on the GaAs nanopillar, the calculated strain magnitude and its distribution as well as the surface and the total energy are considered. We first examine the aspect ratio of an InAs island centered to the nanopillar axis. For a constant number of In and As atoms deposited on the nanopillars, the aspect ratio can be represented by the island incircle radius (Fig. 1). Figure 3 depicts the total energy of the system together with the evaluated strain $\varepsilon$ in [$11\bar{2}$] direction and the number of broken bonds of surface atoms as a function of island incircle radius for a GaAs nanopillar diameter of 25 nm. In order to accommodate all atoms in an island, the InAs (111)A top surface is either atomically flat or it contains a surface step. The energy is represented as the deviation $\Delta E_j$ from that of a reference state $j$, characterized by a flat, step-free InAs (111)A top surface. Since a flat top surface occurs for specific numbers of In and As atoms and incircle radius values, two such reference states $j = 1, 2$ differing only slightly in the number of atoms (by ~3%) are considered exemplarily, in order to reveal the effect of a (111) island surface step on the energetics. When varying the island radius while keeping the number of atoms constant, the length of the surface step shows an oscillatory behavior as function of island radius with maxima for half-filled topmost (111) layers and minima for maximal filling of this layer. In the case of $j = 1$, the island has an incircle radius of ~4.8 nm and a height of 8 monolayers (Fig. 3(a), red-brown arrow), and for $j = 2$ the radius amounts to ~6.2 nm and the height 5 monolayers (Fig. 3(a), black arrow). For the island series pertaining to the first reference state (red-brown diamond dataset in Fig. 3(a)) an energy minimum appears at an island incircle radius of ~4.8 nm (~24% coverage of pillar top face), which results from the combined effects of moderate strain as well as a limited number of broken bonds of atoms on the InAs surfaces and the GaAs top surface (Fig. 3(a) and 3(b)). Certainly, the flat, step-free InAs surface is accompanied by a lower number of broken bonds than a stepped surface leading to an additional reduction of the total energy. In the case of reference state 2, the energy minimum occurs at an island incircle radius of ~6.2 nm (~41% coverage of pillar top face), the maximum for which the island corners are still inside or at the GaAs pillar circumference, corresponding to an aspect ratio of ~0.13 (black squares data set in Fig. 3(a)). Overall, $\Delta E_j$ is low in both datasets for larger island radii corresponding to a pillar top face coverage between 24% and 41%, in qualitative agreement with the experiment. However, the datasets belonging to the two reference states are similar in that the islands with the reference dimensions have particularly low energies and are different in that the number of atoms in InAs is not the same, leading to different lengths of the surface step and related step energies, and thus to different $\Delta E_j$ for $j = 1$ as compared to $j = 2$. Due to the relatively small variations of $\Delta E_j$ for larger island radii, the energy of surface steps has a significant impact on the minimum energy island aspect ratio. Figure 3(c) plots the strain distribution in case of the largest and smallest aspect ratios marked as A and B, respectively, in Fig. 3(a) and 3(b), and depicted in Fig. 3(d). It can be seen that the increased contact area between GaAs and InAs for larger island diameter entails larger strained volumes in GaAs and InAs, and a larger strain magnitude in the regions close to the heterointerface.

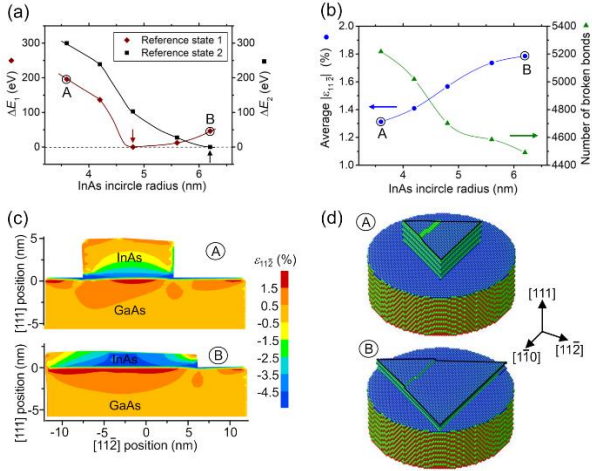

FIG. 3. (a) Deviation of total energy $\Delta E_j$ of a system composed of an InAs island on a 25-nm-diameter GaAs nanopillar from that of the reference state $j$ for two different reference states (marked by an arrow, respectively), (b) averaged $\varepsilon_{11\bar{2}}$ strain magnitude [23] and number of broken bonds of atoms at the InAs surfaces and the GaAs top surface (for reference state 1), as a function of InAs incircle radius, respectively. In (a) and (b) data points are connected with lines to guide the eye. (c) Maps of the $\varepsilon_{11\bar{2}}$ normal strain for two selected island dimensions A, B in the plane defined by [111] and [11$\bar{2}$] directions containing the pillar axis (for reference state 1). (d) Atomistic representations (orthographic projections) of island states A and B. For reasons of space, the GaAs nanopillar has been cut. Island top edges are traced by black lines.

As described by continuum elasticity in Ref. [6], the energy minimum for an island not completely covering the pillar top surface arises because for smaller or larger island in-plane areas, either the surface energy or the strain energy strongly increase, increasing the total energy in both cases. For small island areas, there is a large number of broken bonds of surface atoms whereby the strain concentrates in a small region near the heterointerface. For large island areas on the contrary, there are fewer broken bonds of surface atoms, but the strained volumes in GaAs and InAs are larger.

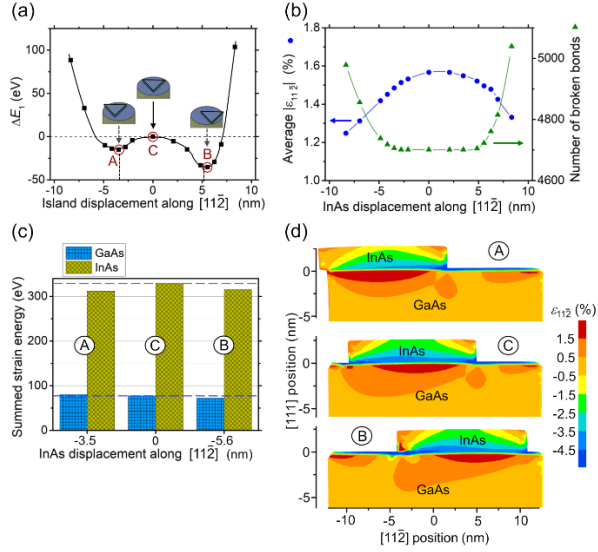

FIG. 4. Energy and strain plotted against the displacement of the InAs island (incircle radius ~4.8 nm) with respect to the GaAs nanopillar along $\pm[11\bar{2}]$: (a) Deviation of the total energy $\Delta E_t$ from the reference state 1 (centered island, solid arrow), dashed vertical lines indicate the positions where the island corners coincide with the pillar circumference. (b) Averaged $\varepsilon_{11\bar{2}}$ strain magnitude [23] and number of broken bonds. In (a) and (b) data points are connected with lines to guide the eye. (c) Comparison of the strain energies due to $\varepsilon_{11\bar{2}}$ of strained GaAs and InAs volumes, for the three island positions A, C, B marked in (a). (d) $\varepsilon_{11\bar{2}}$ strain maps for the three island positions marked in (a) shown in a central cut through pillar and island. The GaAs pillar diameter is 25 nm.

Now let us consider off-axis InAs islands, i.e. islands that are displaced along $\pm[11\bar{2}]$ directions towards the edge of the GaAs nanopillar top face (Fig. 4). The calculations presented here are for an aspect ratio of ~0.27 corresponding to an island incircle radius of ~4.8 nm, i.e. the minimum energy morphology of reference state 1 as discussed above. The total energy plotted against the displacement shows an asymmetric curve with one minimum on each side of the center position. For small displacements with the island corners inside the edge of the nanopillar top face, the number of broken bonds remains constant while the average strain magnitude $|\varepsilon_{11\bar{2}}|$ slightly decreases, lowering the total energy of the system (Fig. 4(a) and 4(b)). As the strain energy continues to decrease for larger displacements, the energy minima are located at positions (A and B in Fig. 4), where one (case A) or two (case B) island corners slightly hang over the pillar edge. If the island displacement exceeds that of positions A and B, the total energy rapidly increases due to the dominating effect of broken bonds of island surface atoms, even though the strain magnitude decreases further. Figure 4(c) compares the strain energies $[Y/(1-\nu)]\sum_i \varepsilon_{11-2,i}^2 \Delta V_i$ of the strained GaAs and InAs volumes for the island center position C and the minimum-energy off-axis positions A and B. $Y$ denotes the Young's modulus, $\nu$ the Poisson ratio, and $\Delta V$ the material volume to which each individual atomistic strain value $\varepsilon_{11\bar{2}}$ is assigned. The summation $i$ runs over all the $\Delta V$ volumes of the respective GaAs or InAs volumes. As a result, the energy benefit of positions A and B originates mainly from a reduced strain energy in InAs due to lower strain in the regions in vicinity of the corners situated above the pillar circumference, which originates from the elastic relaxation at free surfaces. This effect can be seen in the strain map of Fig. 4(d) (A), whereas it is not visible in Fig. 4(d) (B) because the concerned corners are outside the section plane of the strain map. With the island at position A the GaAs lattice is more heavily strained than for positions C and B, since for A the center of the island sits off-axis but still at a distance to the pillar edge. For a centered island GaAs is less deformed because of the relatively large distances to the surfaces, and for an island with the center close to the pillar edge (position B), the elastic relaxation at the GaAs surface leads to a more rapid strain decay with increasing distance from the heterointerface. In total, GaAs and InAs are less strained for the island at position B, giving rise to a more pronounced minimum at position B.

In agreement with the calculations, the island edge-at-pillar-edge configuration (Fig. 4, position B) frequently occurs in the experiment (Fig. 2(a)). The appearance of islands at other positions can be attributed to the relatively small energy differences between different arrangements (Fig. 3(c)) and the possible presence of small surface irregularities of the GaAs nanopillars, such as kinks or nanoscale pits, which lower the energy for island formation at these sites.

## V. CONCLUSIONS

In summary, we find that InAs molecular beam heteroepitaxy at low temperatures on nanopillar-patterned GaAs substrates results in the formation of mostly single InAs islands on the pillar top surfaces, with an island diameter smaller than that of the pillar. Inline with earlier continuum elasticity calculations (for a lattice mismatch of 2%–4%), Tersoff potential based atomistic molecular statics simulations confirm the existence of a minimum-energy state for an intermediate island diameter as a result of the trade-off between strain and surface energy in the case of InAs on GaAs with a higher lattice mismatch of ~7%. Most importantly, we demonstrate that island positions displaced along $\langle 11\bar{2} \rangle$ on the pillar tops, where one side of the triangular island matches the pillar sidewall, are energetically favored compared to the on-axis position owing to the more efficient elastic lattice relaxation of both, GaAs and InAs. This is corroborated by the experiment where such off-axis island positions are often observed. In conclusion, the unveiled energetic preference for heteroepitaxial island nucleation at radially displaced positions on top of nanopillars as a consequence of enhanced elastic relaxation of misfit strains at proximal surfaces becomes important when aiming at the growth of position-controlled, highly mismatched heteroepitaxial islands on top of nanopillars by using catalyst-free vapor phase epitaxy.

## ACKNOWLEDGMENT

The authors are grateful to Deutsche Forschungsgemeinschaft for financial support under Grants No. RI 2655/1-1 and No. LI 449/16-1.

[1] Z. Zhong, A. Halilovic, T. Fromherz, F. Schäffler, and G. Bauer, Two-dimensional periodic positioning of self-assembled Ge islands on prepatterned Si (001) substrates, Appl. Phys. Lett. **82**, 4779 (2003).

[2] O. Skibitzki, I. Prieto, R. Kozak, G. Capellini, P. Zaumseil, Y. A. R. Dasilva, M. D. Rossell, R. Erni, H. von Känel, and T. Schroeder, Structural and optical characterization of GaAs nano-crystals selectively grown on Si nano-tips by MOVPE, Nanotechnol. **28**, 135301 (2017).

[3] M. T. Björk, B. J. Ohlsson, C. Thelander, A. I. Persson, K. Deppert, L. R. Wallenberg, and L. Samuelson, Nanowire resonant tunneling diodes, Appl. Phys. Lett. **81**, 4458 (2002).

[4] E. D. Minot, F. Kelkensberg, M. van Kouwen, J. A. van Dam, L. P. Kouwenhoven, V. Zwiller, M. T. Borgström, O. Wunnicke, M. A. Verheijen, and E. P. A. M. Bakkers, Single Quantum Dot Nanowire LEDs, Nano Lett. **7**, 367 (2007).

[5] M. T. Borgström, V. Zwiller, E. Müller, and A. Imamoglu, Optically Bright Quantum Dots in Single Nanowires, Nano Lett. **5**, 1439 (2005).

[6] F. Glas and B. Daudin, Stress-driven island growth on top of nanowires, Phys. Rev. B **86**, 174112 (2012).

[7] G. Tourbot, C. Bougerol, F. Glas, L. F. Zagonel, Z. Mahfoud, S. Meuret, P. Gilet, M. Kociak, B. Gayral, and B. Daudin, Growth mechanism and properties of InGaN insertions in GaN nanowires, Nanotechnology **23**, 135703 (2012).

[8] M. Morassi, L. Largeau, F. Oehler, H.-G. Song, L. Travers, F. H. Julien, J.-C. Harmand, Y.-H. Cho, F. Glas, M. Tchernycheva, and N. Gogneau, Morphology Tailoring and Growth Mechanism of Indium-Rich InGaN/GaN Axial Heterostructures by Plasma-Assisted Molecular Beam Epitaxy, Cryst. Growth Des. **18**, 2545 (2018).

[9] J. Bartolomé, M. Hanke, D. van Treeck, and A. Trampert, Strain Driven Shape Evolution of Stacked (In,Ga)N Quantum Disks Embedded in GaN Nanowires, Nano Lett. **17**, 4654 (2017).

[10] B. Yang, F. Liu, and M. G. Lagally, Local Strain-Mediated Chemical Potential Control of Quantum Dot Self-Organization in Heteroepitaxy, Phys. Rev. Lett. **92**, 025502 (2004).

[11] X. L. Li, G. Ouyang, and G. W. Yang, Thermodynamic theory of nucleation and shape transition of strained quantum dots, Phys. Rev. B **75**, 245428 (2007).

[12] Z.-R. Lv, Z.-K. Zhang, X.-G. Yang, and T. Yang, Improved performance of 1.3-μm InAs/GaAs quantum dot lasers by direct Si doping, Appl. Phys. Lett. **113**, 011105 (2018).

[13] H. Yoshikawa, J. Kwoen, T. Doe, M. Izumi, S. Iwamoto, and Y. Arakawa, InAs/GaAs quantum dot infrared photodetectors on on-axis Si (100) substrates, Electronics Lett. **54**, 1395 (2018).

[14] M. V. Rakhlin, K. G. Belyaev, G. V. Klimko, I. S. Mukhin, D. A. Kirilenko, T. V. Shubina, S. V. Ivanov, and A. A. Toropov, InAs/AlGaAs quantum dots for single-photon emission in a red spectral range, Sci. Rep. **8**, 5299 (2018).

[15] H. Xie, R. Prioli, A. M. Fischer, F. A. Ponce, R. M. S. Kawabata, L. D. Pinto, R. Jakomin, M. P. Pires, and P. L. Souza, Improved optical properties of InAs quantum dots for intermediate band solar cells by suppression of misfit strain relaxation, J. Appl. Phys. **120**, 034301 (2016).

[16] V. S. Kunnathully, T. Riedl, A. Trapp, T. Langer, D. Reuter, and J. K. N. Lindner, InAs heteroepitaxy on nanopillar-patterned GaAs (111)A, arXiv:1909.08480.

[17] J. Tersoff, New empirical approach for the structure and energy of covalent systems, Phys. Rev. B **37**, 6991 (2016).

[18] Gatan Inc., Gatan Microscopy Suite Software. http://www.gatan.com/products/tem-analysis/gatan-microscopy-suite-software (accessed 19 September, 2019).

[19] S. Plimpton, Fast Parallel Algorithms for Short-Range Molecular Dynamics, J. Comp. Phys. **117**, 1 (1995).

[20] T. Hammerschmidt, P. Kratzer, and M. Scheffler, Analytic many-body potential for InAs/GaAs surfaces and nanostructures: Formation energy of InAs quantum dots, Phys. Rev. B **77**, 235303 (2008).

[21] J. S. Blakemore, Semiconducting and other major properties of gallium arsenide, J. Appl. Phys. **53**, R123 (1982).

[22] Yu. A. Burenkov, S. Yu. Davydov, and S. P. Nikanorov, Elastic properties of indium arsenide, Sov. Phys. Solid State **17**, 1446 (1975).

[23] Averaged over GaAs regions within ~2 nm distance from the GaAs top surface and InAs.